\documentclass[preprint,showpacs,preprintnumbers,superscriptaddress,amsmath,amssymb]{revtex4-1}
\usepackage[dvipdfmx]{graphicx}
\usepackage{colortbl}
\usepackage{mathrsfs}
\usepackage{lineno}

\usepackage{bm}  
\usepackage{ulem}
\usepackage{amsmath}

\renewcommand{\Im}{\operatorname{Im}}

\makeatletter
\def\btt#1{\texttt{\@backslashchar#1}}%
\DeclareRobustCommand\bblash{\btt{\@backslashchar}}%
\makeatother

\begin{document}


\title{Symmetry of the emergent inductance tensor exhibited by magnetic textures}

\author{Soju Furuta}
\affiliation{Department of Physics, Tokyo Institute of Technology, Tokyo 152-8551, Japan}

\author{Wataru Koshibae}
\affiliation{RIKEN Center for Emergent Matter Science (CEMS), Wako 351-0198, Japan}

\author{Fumitaka Kagawa}
\email{kagawa@phys.titech.ac.jp}
\affiliation{Department of Physics, Tokyo Institute of Technology, Tokyo 152-8551, Japan}
\affiliation{RIKEN Center for Emergent Matter Science (CEMS), Wako 351-0198, Japan}

\date{\today}
\begin{abstract}
Metals hosting gradually varying spatial magnetic textures are attracting attention as a new class of inductor. Under the application of an alternating current, the spin-transfer-torque effect induces oscillating dynamics of the magnetic texture, which subsequently yields the spin-motive force as a back action, resulting in an inductive voltage response. In general, a second-order tensor representing a material's response can have an off-diagonal component. However, it is unclear what symmetries the emergent inductance tensor has and also which magnetic textures can exhibit a transverse inductance response. Here we reveal both analytically and numerically that the emergent inductance tensor should be a symmetric tensor in the so-called adiabatic limit. By considering this symmetric tensor in terms of symmetry operations that a magnetic texture has, we further characterize the magnetic textures in which the transverse inductance response can appear. This finding provides a basis for exploring the transverse response of emergent inductors, which has yet to be discovered.
\end{abstract}

\maketitle

\section*{I\lowercase{ntroduction}}

An inductor is a component that exhibits an inductive counter-electromotive force, $V$, under a time-varying electric current, $I$, following
\begin{equation}
\label{L}
V = L\frac{{\rm d}I}{{\rm d}t},
\end{equation}
where $L$ denotes the inductance. The electric work done by the external power supply, $IV$, is hence 
\begin{equation}
\label{energy}
\int {\rm d}t \: IV = \int {\rm d}t \: IL\frac{{\rm d}I}{{\rm d}t} = \int {\rm d} \left( \frac{1}{2}LI(t)^2 \right),
\end{equation}
which shows that the inductor stores an energy of $\frac{1}{2}LI^2$. Thus, it can also be said that an inductor is a component that can store an energy of $\Delta E= \frac{1}{2}LI^2$, under the application of an electric current. A textbook example is a solenoidal inductor, which stores energy as a magnetic-field energy \cite{Jackson}. Other inductors possess similar energy-storing properties. An established example is the so-called kinetic inductor, in which the energy is stored as the kinetic energy of mobile charge carriers. When considering the Drude model of conduction electrons, one can immediately find that the inductance defined using the imaginary part of the angular-frequency ($\omega$)-dependent resistivity, $\rho(\omega)$, agrees with the inductance defined using the total kinetic energy of electrons \cite{kinetic}.

Recently, a new class of inductor, now referred to as emergent inductors, has been proposed theoretically \cite{NagaosaJJAP} and confirmed experimentally \cite{YokouchiNature, KitaoriPNAS, KitaoriPRB}. In these inductors, the flowing conduction electrons exert a spin-transfer torque (STT) \cite{STT1, STT2, YamanouchiNature, YamaguchiPRL} on the underlying magnetic texture; as a result, the magnetic texture exhibits time-dependent elastic deformations under an alternating current (AC) in the linear-response regime. Such current-induced magnetic texture dynamics exert a back action on the flowing conduction electrons, yielding the so-called spin-motive force or emergent electric field (EEF) \cite{Volovik, BarnesPRL2007, YangPRL, JiadongPRL, Schulz_Nat.Phys.}. This phenomenon can be derived microscopically in terms of the so-called spin-Berry phase or the effective U(1) gauge field, and the resulting EEF can be described by
\begin{equation}
\label{e-field}
e_i(\bm{r}, t) = \frac{\hbar}{2|e|} \bm{m}(\bm{r}, t) \cdot [{\partial}_i \bm{m}(\bm{r}, t) \times {\partial}_t \bm{m}(\bm{r}, t)],
\end{equation}
where $e$ $(>0)$ is the elementary charge, $\bm{m}(\bm{r}, t)$ is the unit vector of the local magnetic moment at position $\bm{r}$ and time $t$, and ${\partial}_i$ ($i = x, y, z$) and ${\partial}_t$ denote spatial and time derivatives, respectively (when the conduction-electron spins are not fully polarized, the so-called spin-polarization factor $P$ is further multiplied on the right-hand side of Eq.~(\ref{e-field}) \cite{Schulz_Nat.Phys., BarnesPRL2007}). It has been numerically demonstrated that in the so-called adiabatic limit (i.e., $\beta = 0$; see the Methods section), the inductance value defined using the EEF under an AC quantitatively agrees with that defined using the current-induced magnetic-texture-deformation energy \cite{Furuta}. Thus, in the adiabatic limit, the emergent inductance is well defined, and both the electric and energetic responses are correctly captured by Eq.~(\ref{L}). On the other hand, when nonadiabaticity is concerned (i.e., $\beta \neq 0$), the inductance values derived independently from the two definitions do not match, implying that the system responses are beyond the framework of Eq.~(\ref{L}) and hence the inductance interpretation does not apply.

An interesting aspect of emergent inductors is that the inductive electric response is potentially not limited to the applied current direction but may also appear along the perpendicular directions, as inferred from Eq.~(\ref{e-field}). Thus, in general, the emergent inductance, when it is well defined, should be represented by a tensor: $V_i = L_{ij}\frac{{\rm d}I_j}{{\rm d}t} (i, j = x, y)$ or
\begin{eqnarray}
\label{tensorL}
\begin{pmatrix}
   V_x \\
   V_y
\end{pmatrix}
=
\begin{pmatrix}
   L_{xx} & L_{xy} \\
   L_{yx} & L_{yy}
\end{pmatrix}
\frac{{\rm d}}{{\rm d}t}
\begin{pmatrix}
   I_x \\
   I_y
\end{pmatrix}
.
\end{eqnarray}
In classical electrodynamics, such an inductance tensor with $i, j = 1, 2$ may be introduced to describe two mutually coupled coils, 1 and 2. It therefore appears that an emergent inductor possesses a function similar to that of a coupled classical inductor system. However, such an intuitive analogy requires careful consideration because the microscopic mechanism is quite different between classical and emergent inductors. For instance, in a coupled classical inductor system, one can analytically express the mutual inductance and find $L_{12} = L_{21} \equiv M$ \cite{Jackson}; moreover, the fact that the coupled system stores a positive energy in the quadratic form of $\frac{1}{2}L_{11}I_1^2 + \frac{1}{2}L_{22}I_2^2 +MI_1I_2$ for arbitrary values of $I_1$ and $I_2$ leads to a constraint, $L_{11}L_{22} \geq M^2$ \cite{Feynman}, in addition to the obvious one, $L_{11}, L_{22} \geq 0$. Such classical electrodynamics considerations, however, are not helpful for emergent inductors consisting of an arbitrary spin texture including disorder, and thus, the relation between $L_{xy}$ and $L_{yx}$ appears to be nontrivial.

When considering the nature of $L_{ij}$, it is instructive to review the resistivity tensor, $\rho_{ij}$, as a textbook example. Note that any second-order tensor, $K_{ij}$, can always be decomposed into a symmetric part, $K_{ij}^{\rm S}$, and an antisymmetric part, $K_{ij}^{\rm A}$; namely, $K_{ij} = K_{ij}^{\rm S} +K_{ij}^{\rm A}$ with $K_{ij}^{\rm S} = (K_{ij}+K_{ji})/2$ and $K_{ij}^{\rm A} = (K_{ij}-K_{ji})/2$. In the case of $\rho_{ij}$, the symmetric part represents dissipative transport, whereas the antisymmetric part represents nondissipative transport, that is, the Hall resistivity. Thus, the symmetric and antisymmetric parts of $\rho_{ij}$ have their own physical meanings with quite different characteristics. Therefore, the symmetry of the emergent inductance tensor is also an important issue in understanding the underlying physics. 

In this paper, focusing on the adiabatic limit, in which the emergent inductance is well defined by Eq.~(\ref{L}) \cite{Furuta}, we aim to reveal the symmetry of the emergent inductance tensor and discuss the physical implications of the revealed symmetry. Our approach is two-fold. First, we consider the tensor-expressed circuit equation [Eq.~(\ref{tensorL})] in detail and draw a conclusion regarding the symmetry of $L_{ij}$: this also enables us to discuss how the inductor tensor should behave under the time-reversal operation. Second, we numerically investigate $L_{ij}$ for various magnetic textures using micromagnetic simulations. These two approaches consistently show that $L_{ij}$ is a symmetric tensor (that is, $L_{xy} = L_{yx}$) and $L_{ij}$ is even under the time-reversal operation. By combining the numerical results and symmetry arguments, we also find what kinds of magnetic textures can or cannot exhibit a transverse emergent inductance, $L_{yx}$. We note that the present conclusion is for the case where the emergent inductance is well defined (i.e., the adiabatic limit, $\beta=0$). The effect of nonadiabaticity (i.e., $\beta \neq 0$), which makes the emergent inductance ill-defined \cite{Furuta}, is discussed in the Supplementary Information.

\section*{R\lowercase{esults}}

\subsection*{C\lowercase{onsiderations for the circuit equation}}

We discuss the consequences that are prescribed in the tensor-expressed circuit equation, Eq.~(\ref{tensorL}). Following the general arguments on a second-order tensor, we decompose $L_{ij}$ into symmetric and antisymmetric parts: $L_{ij} = L_{ij}^{\rm S} +L_{ij}^{\rm A}$, or explicitly,
\begin{equation}
\begin{pmatrix}
   L_{xx} & L_{xy} \\
   L_{yx} & L_{yy}
\end{pmatrix}
=
\begin{pmatrix}
   L_{xx}^{\rm S} & L_{xy}^{\rm S} \\
   L_{xy}^{\rm S} & L_{yy}^{\rm S}
\end{pmatrix}
\\
+
\begin{pmatrix}
   0 & L_{xy}^{\rm A} \\
   -L_{xy}^{\rm A} & 0
\end{pmatrix}
.
\end{equation}
To gain insight into the physical meaning of the symmetric and antisymmetric tensors, we consider the work done by the power source along a closed loop in the $I_x$--$I_y$ plane, which is expressed as:
\begin{eqnarray}
\label{work}
\begin{split}
\oint{\rm d}t \: (I_xV_x + I_yV_y) = \oint{\rm d} \left( \frac{1}{2}L_{xx}^{\rm S}I_x^2 + \frac{1}{2}L_{yy}^{\rm S}I_y^2 +L_{xy}^{\rm S}I_xI_y  \right) \\
+ \oint{\rm d}t \: L_{xy}^{\rm A}\left( I_x\frac{{\rm d}}{{\rm d}t}I_y - I_y\frac{{\rm d}}{{\rm d}t}I_x  \right).
\end{split}
\end{eqnarray}
Note that the first term in the right-hand side consists of only the symmetric tensor components and the integrand takes the form of a total derivative; hence, the contour integral results in zero. The expression, $\frac{1}{2}L_{xx}^{\rm S}I_x^2 + \frac{1}{2}L_{yy}^{\rm S}I_y^2 +L_{xy}^{\rm S}I_xI_y$, is essentially the same as that derived for mutually coupled classical inductors, representing an energy stored in the emergent inductor under a current. In contrast, the second term consists of only the antisymmetric-tensor components, and the integrand is not the form of a total derivative, indicating that the second term is nonzero and dependent on the path. These features imply that $L_{xy}^{\rm A}$ is associated with a non-conserved quantity.

\begin{figure}
\includegraphics{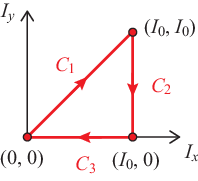}
\caption{\label{} \textbf{A specific closed loop used to prove the absence of the antisymmetric components of an inductance tensor.}}
\end{figure}

To see the consequences of the antisymmetric component $L_{xy}^{\rm A}$ more clearly, it is helpful to consider a specific closed path for the integral(s) of Eq.~(\ref{work}). Suppose $L_{xy}^{\rm A} > 0$; we consider a specific cycle $C$ that consists of three paths, $C_1$, $C_2$ and $C_3$, as shown in Fig.~1: $(I_x, I_y) = (0, 0) \xrightarrow{C_1} (I_0, I_0) \xrightarrow{C_2} (I_0, 0) \xrightarrow{C_3} (0, 0)$ with constraints of $I_x=I_y$ on $C_1$, $I_x=I_0$ on $C_2$, and $I_y=0$ on $C_3$. Thus, taking the contour integral along the cycle in the clockwise direction results in:
\begin{equation}
\oint_C{\rm d}t \: (I_xV_x + I_yV_y) = - L_{xy}^{\rm A}I_0^2 < 0.
\end{equation}
The result indicates that if a positive $L_{xy}^{\rm A}$ were present, the power source could acquire energy by cycling the closed loop. Such behaviour is obviously not allowed for a passive element, such as a stable material. Similarly, one can consider the case of $L_{xy}^{\rm A} < 0$, and the same conclusion can be drawn by considering the same closed loop $C$ but in the counterclockwise direction. Thus, Eq.~(\ref{tensorL}) concludes that even for the case of an emergent inductor, the inductance tensor cannot have an antisymmetric component; that is, $L_{xy}^{\rm A} = 0$, and an emergent inductor tensor should be a symmetric tensor (below, we therefore omit the superscript, S),
\begin{equation}
L_{xy} = L_{yx} \equiv L_{tr}.
\end{equation}
Hence, the energy stored in an emergent inductor under current is found to be expressed by $\frac{1}{2}L_{xx}I_x^2 + \frac{1}{2}L_{yy}I_y^2 +L_{tr}I_xI_y$, and for this quadratic form to be nonnegative, $L_{ij}$ should satisfy
\begin{equation}
L_{xx}L_{yy} \geq (L_{tr})^2,
\end{equation}
in addition to $L_{xx}, L_{yy} \geq 0$ \cite{Furuta}.

Thus, although the microscopic mechanism is quite different between classical and emergent inductors, it turns out that there is no difference in the constraints that inductance tensors should satisfy. These characteristics are implicitly prescribed by the relation between voltage and current, Eq.~(\ref{tensorL}), not depending on the microscopic mechanism for inductors.

Having established the symmetry of $L_{ij}$, we can discuss the behaviour of $L_{ij}$ under the time-reversal operation. Since an emergent inductance arises from a magnetic texture $\{ \bm{m}(\bm{r}) \}$, the behaviour of an emergent inductance under the time-reversal operation is an interesting issue. In fact, in experiments, the magnetic-field ($\bm{B}$)-dependence of an emergent inductance has been frequently investigated \cite{YokouchiNature, KitaoriPNAS, KitaoriPRB}. To incorporate a case where $\{ \bm{m}(\bm{r}) \}$ shows hysteretic behaviour with respect to changes in $\bm{B}$, $L_{ij}$ may be expressed as a function with $\{ \bm{m}(\bm{r}) \}$ and $\bm{B}$ as variables. Note that regardless of the details of the variables, $L_{ij}$ should be a symmetric tensor as discussed above, and hence, $L_{ij}(\bm{B}, \{ \bm{m}(\bm{r}) \}) = L_{ji}(\bm{B}, \{ \bm{m}(\bm{r}) \})$ should always be satisfied. Moreover, with respect to the complex resistivity, Onsager's reciprocal theorem concludes $\Im \rho_{ij}(\omega, \bm{B}, \{ \bm{m}(\bm{r}) \}) = \Im \rho_{ji}(\omega, -\bm{B}, \{ -\bm{m}(\bm{r}) \})$ (the real part also satisfies the same relation) \cite{Kubo}; hence, the inductance tensor should also satisfy $L_{ji}(\bm{B}, \{ \bm{m}(\bm{r}) \}) = L_{ij}(-\bm{B}, \{ -\bm{m}(\bm{r}) \})$. 
By combining the two relations regarding $L_{ij}$, one can thus conclude
\begin{equation}
L_{ij}(\bm{B}, \{ \bm{m}(\bm{r}) \}) = L_{ij}(-\bm{B}, \{ -\bm{m}(\bm{r}) \}).
\end{equation}
This relation indicates that $L_{ij}$ is even under time-reversal, or equivalently, $L_{ij}$ is a polar symmetric tensor. In particular, we note $L_{yx}(\bm{B}, \{ \bm{m}(\bm{r}) \}) = L_{yx}(-\bm{B}, \{ -\bm{m}(\bm{r}) \})$, distinct from the Hall resistivity, which satisfies $\rho_{yx}(\bm{B}, \{ \bm{m}(\bm{r}) \}) = - \rho_{yx} (-\bm{B}, \{ -\bm{m}(\bm{r}) \})$. For this reason, we call $L_{yx}$ the transverse inductance, not the Hall inductance.

\subsection*{M\lowercase{icromagnetic simulations}}

To observe the symmetry of the inductance tensor of emergent inductors, we consider magnetic textures that slowly vary in space; for such magnetic textures, the EEF can be calculated according to Eq.~(\ref{e-field}). We further consider the pinned regime, in which a magnetic texture does not exhibit a steady flow under a direct current \cite{Nattermann, ChauvePRB, Kleemann, IntrinsicTatara, ThiavilleEPL, ExtrinsicTatara, IntrinsicOno, TataraReview, ExtrinsicNatPhys, Iwasaki_Nat.Commun.}. The procedure for calculating the emergent inductance arising from a slowly varying magnetic texture in the pinned regime is detailed in the literature \cite{Furuta} and also in the Methods section. We consider a spin Hamiltonian based on the continuum approximation that can exhibit helical and skyrmion-lattice (SkL) \cite{Bogdanov1, Bogdanov2, Muhlbauer_Science, Yu_Nature} magnetic textures and calculate the current-induced dynamics of a magnetic texture by numerically solving the Landau-Lifshitz-Gilbert (LLG) equation \cite{LLG} (see the Methods section). To be more specific, the magnetic texture dynamics under the application of an AC along the $x$-direction are calculated by micromagnetic simulation; then, by referring to Eq.~(\ref{e-field}), the time-dependent EEFs along both the $x$ and $y$ directions are further derived; and finally, by referring to Eq.~(\ref{tensorL}), $L_{xx}$ and $L_{yx}$ are obtained. Similarly, we obtain $L_{yy}$ and $L_{xy}$ by simulating the case of an AC along the $y$ direction. The emergent inductance $L_{ij}$ depends on the system dimension in the form of $L_{ij} = \tilde{L}_{ij}\frac{\ell}{S}$, where $\tilde{L}_{ij}, \ell$, and $S$ represent the normalized inductance (we call it ``inductivity''), system length, and sample cross-section area. Below, we therefore present $\tilde{L}_{ij}$, rather than the system-size-dependent $L_{ij}$. The inductivity tensor may be defined by
\begin{equation}
\label{L_tilde}
e_i = \tilde{L}_{ik}\frac{{\rm d}j_k}{{\rm d}t},
\end{equation}
where $j$ represents the current density. The following simulation results are obtained for the case of $\beta = 0$ (i.e., the adiabatic limit).

\begin{figure*}
\includegraphics{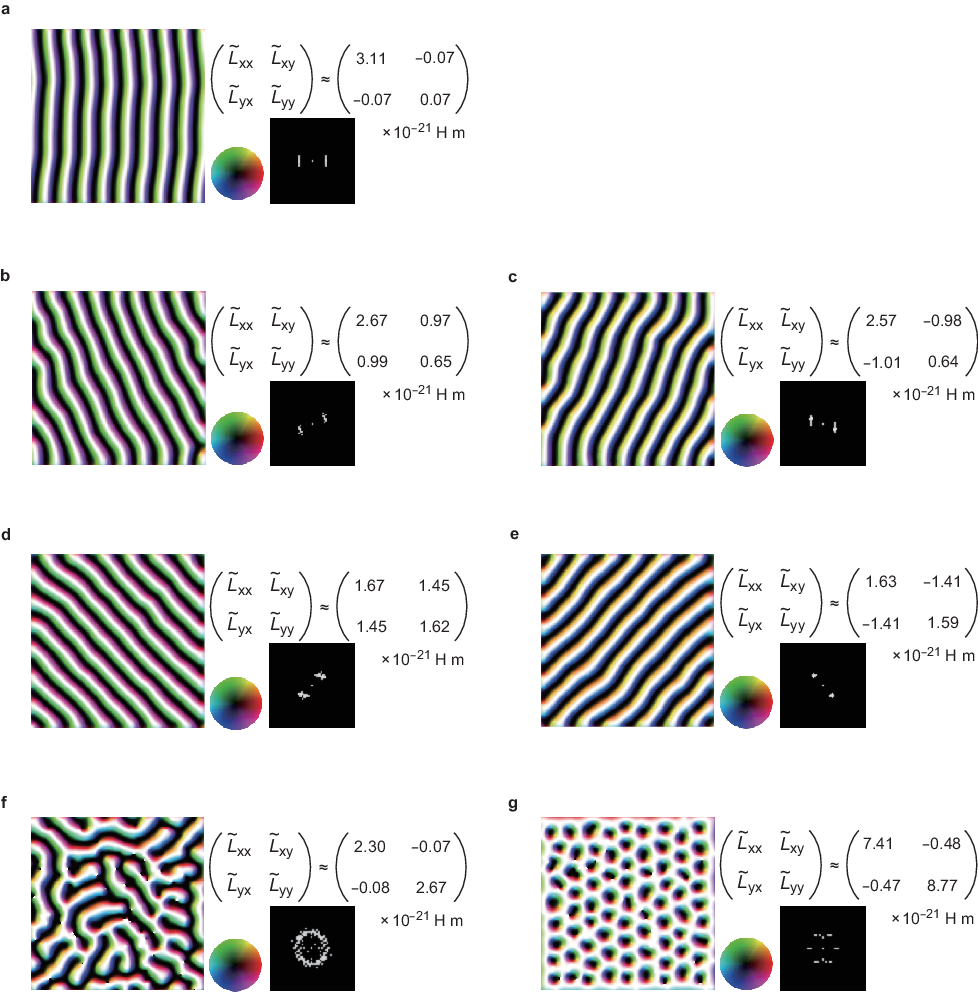}
\caption{\label{} \textbf{Various metastable magnetic textures and corresponding inductance tensors.} \textbf{a-e} Helical magnetic textures with the helical ${\bm q}$-vector that forms approximately an angle $\theta = 0^{\circ}$ (\textbf{a}), $20^{\circ}$ (\textbf{b}), $-20^{\circ}$ (\textbf{c}), $45^{\circ}$ (\textbf{d}), and $-45^{\circ}$ (\textbf{e}) with respect to the $x$ direction. \textbf{f} Maze helix. \textbf{g} Skyrmion lattice. The corresponding fast-Fourier-transform (FFT) images are also shown in each panel. Color wheels specify the $x$-$y$ plane magnetization direction. The brightness of the color represents the $z$ component of the magnetization, and white represents the local magnetizations pointing toward the $z$ direction. The current-induced magnetic texture dynamics are calculated under the application of a weak AC. The parameters used for the simulation are tabulated in Table I; they are chosen so that the resulting emergent voltage is in the linear-response and low-frequency regimes (see the Methods section). The simulations were done for $\beta = 0$.}
\end{figure*}

Figure 2 summarizes the magnetic textures investigated in this study and the corresponding inductance tensors. We studied four examples of helical magnetic textures, for which the helical ${\bm q}$-vector forms approximately an angle $\theta = 0^{\circ}, \pm 20^{\circ}$, and $\pm 45^{\circ}$ with respect to the $x$ direction (Fig.~2a--e, respectively); a maze-helix texture (Fig.~2f); and an SkL (Fig.~2g). The intensity and concentration of disorder were minimized as much as possible while confirming the linear response of the pinned dynamics. As a result, more disorder had to be included when examining the maze-helix and SkL, as summarized in Table I: Selecting a much lower current density while keeping the disorder density as low as 0.3 $\%$ was not appropriate in terms of the required numerical accuracy. As shown in Fig.~2, we find that $\tilde{L}_{xy} = \tilde{L}_{yx}$ invariably holds within the numerical error, consistent with the conclusion derived from the circuit equations. 

\begin{table*}[t]
 \caption{Parameters used for the micromagnetic simulations displayed in Fig.~2. The disorder density was chosen as low as possible while confirming the linear response of the pinned magnetic textures under a given current density.
}
  \centering
  \begin{tabular}{cccccc}
   \hline
   Magnetic texture & Magnetic field & $K_{\rm imp}$ & Disorder density & Current density & Frequency\\
    & (T) & $\times 10^{7}$ (J m$^{-3}$) & ($\%$) & $\times 10^{10}$ (A m$^{-2}$) & (MHz) \\
   \hline \hline
   Helix ($\theta =0, \pm 20^{\circ}, \pm 45^{\circ}$) & 0 & 0.1 & 0.3 & 5.0 & 50 \\
   Maze helix & 0 & 2.0 & 3 & 2.0 & 50 \\
   SkL & 0.3 & 1.0 & 3 & 1.0 & 10 \\
   \hline
  \end{tabular}
\end{table*}

\begin{figure}
\includegraphics{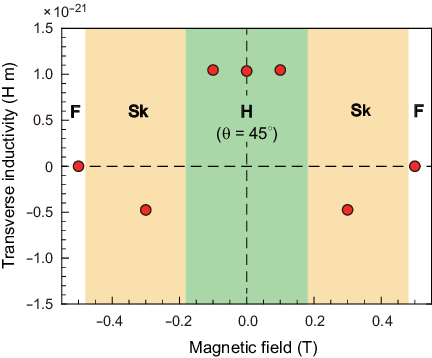}
\caption{\label{} \textbf{Magnetic-field dependence of the transverse inductivity.} H, Sk, and F denote the single ${\bm q}$-helix with $\theta = 45^{\circ}$, skyrmion lattice, and ferromagnetic state, respectively. The disorder density and strength $K_{\rm imp}$ (see Methods) is fixed to 3 $\%$ and 1.0$\times$10$^{7}$ J m$^{-3}$. The simulations were done for $\beta = 0$. When $\beta$ is non-zero, the effective inductivity tensor $\tilde{L}^{\rm eff}_{ij}$ should be discussed ($\tilde{L}^{\rm eff}_{ij} = \Im [\rho_{ij}(\omega)- \rho_{ij}(0)]/\omega$): The magnetic-field dependence of the transverse components of $\tilde{L}^{\rm eff}_{ij}$ at finite $\beta$ is shown in Fig.~S1 and discussed in Supplementary Note 2. Note that $\tilde{L}^{\rm eff}_{ij}$ is a different quantity from the inductivity tensor defined by Eq.~(11).}
\end{figure}

The diagonal components of the inductance tensor are invariably positive, whereas the off-diagonal components can be either positive or negative. Nevertheless, we emphasize that the inductance tensor retains energetic interpretations; that is, the energy increase in the magnetic system, $\Delta E(I_x, I_y)$, caused by the application of an electric current agrees with $\frac{1}{2}L_{xx}I_{x}^2 Fig.~3+ \frac{1}{2}L_{yy}I_{y}^2 + L_{tr}I_{x}I_{y}$. As an example, we discuss the results for the helical texture with $\theta = -20^{\circ}$, in which the off-diagonal components of $\tilde{L}_{ij}$ are negative. The $\Delta E(I_x, I_y)$ is calculated for the following three cases independently: (i) $(I_x \neq 0, I_y = 0)$, (ii) $(I_x = 0, I_y \neq 0)$, and (iii) $(I_x \neq 0, I_y \neq 0)$. Then, by solving the three simultaneous equations regarding $\Delta E(I_x, I_y) = \frac{1}{2}L_{xx}I_{x}^2 + \frac{1}{2}L_{yy}I_{y}^2 + L_{tr}I_{x}I_{y}$, we can obtain: $(\tilde{L}_{xx}, \tilde{L}_{yy}, \tilde{L}_{tr}) = (2.67, 0.63, -1.00) \times10^{-21}$ H~m. These values are in quantitative agreement with $\tilde{L}_{ij}$ calculated from the EEF (Fig.~2c), indicating that the emergent inductivity is well defined by Eq.~(\ref{L_tilde}). We also confirmed $\tilde{L}_{xy}(\bm{B}, {\bm{m}(\bm{r}_k)}) = \tilde{L}_{xy}(-\bm{B}, {-\bm{m}(\bm{r}_k)})$ numerically (Fig.~3), in which the disorder density and strength $K_{\rm imp}$ (see Methods) are fixed to 3 $\%$ and 1.0$\times$10$^{7}$ J m$^{-3}$, respectively, and the single-${\bm q}$ helix with $\theta = 45^{\circ}$ was considered. Thus, our numerical study confirms that $\tilde{L}_{ij}$ for an emergent inductor is a polar symmetric tensor.

When comparing the three helical textures quantitatively, one can find that as the $\theta$ increases from 0$^{\circ}$ to 45$^{\circ}$, $\tilde{L}_{xx}$ decreases, whereas $\tilde{L}_{xy}$ increases. We also note that in the maze-helix and SkL textures, the transverse component, $\tilde{L}_{xy}$, is more than one order of magnitude smaller than the longitudinal components, $\tilde{L}_{xx}$ and $\tilde{L}_{yy}$. As discussed below, these observations can be explained by considering an orthogonal transformation of $\tilde{L}_{ij}$ and the rotational symmetry that each magnetic texture has.

\section*{D\lowercase{iscussion}}

In the following, we aim to categorize inductance tensors of a magnetic texture origin and consider how our numerical results obtained in the adiabatic limit can be explained in terms of the symmetry operations that each magnetic system has. Note that because an inductance tensor is real and symmetric, it can be diagonalized by performing an appropriate orthogonal transformation, $R$, or equivalently by choosing appropriate Cartesian coordinates: 
\begin{equation}
\label{Diagonalization}
\begin{pmatrix}
   \tilde{L}_{xx} & \tilde{L}_{tr} \\
   \tilde{L}_{tr} & \tilde{L}_{yy}
\end{pmatrix}
\xrightarrow{R}
\begin{pmatrix}
   \lambda_1 & 0 \\
   0 & \lambda_2
\end{pmatrix}
,
\end{equation}
where $\lambda_1$ and $\lambda_2$ ($\lambda_1, \lambda_2 \geq 0$) represent the eigenvalues of the inductance tensor. Hence, to classify an emergent inductance tensor, it is sufficient to consider the diagonalized form. This approach does not lose generality because a representation in different Cartesian coordinates can be immediately obtained by performing the corresponding orthogonal transformation. Following the group theory arguments for a polar symmetric tensor, one can conclude that: (i) when the system has three-fold or higher rotational symmetry with respect to the $z$ axis (i.e., $C_{3z}, C_{4z}, C_{6z}$ or $C_{\infty z}$), $\lambda_1$ and $\lambda_2$ should be equal, whereas (ii) when the system has only two-fold with respect to the $z$ axis ($C_{2z}$) or no rotational symmetry, $\lambda_1$ and $\lambda_2$ should be inequivalent: For the details of the derivation, see Supplementary Note 3. Thus, the diagonalized two-by-two tensor can be classified as one of the two categories, which are characterized by $\lambda_1 = \lambda_2$ and $\lambda_1 \neq \lambda_2$, respectively.

The first category, $\lambda_1 = \lambda_2$, is represented by an isotropic tensor $\begin{pmatrix}
   1 & 0 \\
   0 & 1
\end{pmatrix}$,
and thus, the off-diagonal components are always zero for arbitrarily chosen Cartesian coordinates; that is, the transverse inductance response does not appear. From group theory, a magnetic texture that has $C_{3z}, C_{4z}$, $C_{6z}$ or $C_{\infty z}$ symmetry should belong to this category. Note that our numerical calculations deal with finite-size systems including randomly distributed disorder, and therefore, the simulated magnetic textures do not have any rotational symmetry in a strict sense. Nevertheless, we numerically find that the inductance tensors of the maze-helix and SkL textures satisfy $\tilde{L}_{xx} \approx \tilde{L}_{yy}$ and $\tilde{L}_{xy}, \tilde{L}_{yx} \ll \tilde{L}_{xx}, \tilde{L}_{yy}$ (Fig.~2f and g, respectively), indicating that the obtained tensors are close to isotropic. These results appear reasonable, considering that in a macroscopic system, the maze-helix and SkL textures have global approximate $C_{\infty z}$ and $C_{6z}$ symmetries, respectively. The symmetry of the macroscopic systems can be imagined by looking at the corresponding fast-Fourier-transform (FFT) images. To be precise, the rotation symmetry of the SkL confined in the finite-size system is $C_{2z}$, rather than $C_{6z}$, as indicated by the FFT image (Fig.~2g): This perturbative symmetry lowering from $C_{6z}$ to $C_{2z}$ explains the small but finite symmetric off-diagonal component, which is originally prohibited under $C_{6z}$ symmetry. When nonadiabaticity is not negligible, the effective inductivity tensor defined by $\tilde{L}^{\rm eff}_{ij} = \Im [\rho_{ij}(\omega)- \rho_{ij}(0)]/\omega$ is discussed, but it should be noted that $\tilde{L}^{\rm eff}_{ij}$ is a different quantity from the inductivity tensor in Eq.~(\ref{L_tilde}). For instance, the $\tilde{L}^{\rm eff}_{ij}$ of the SkL has antisymmetric off-diagonal components when $\beta \neq 0$, although the antisymmetric component in $\tilde{L}_{ij}$ is energetically prohibited; for more details, see Supplementary Notes 1 and 3.

\begin{figure}
\includegraphics{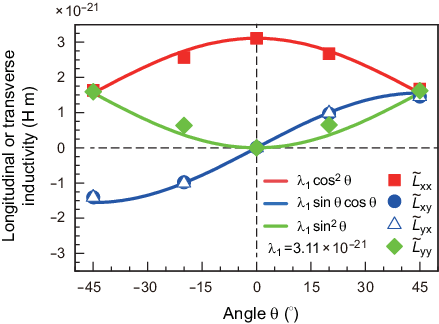}
\caption{\label{} \textbf{Numerically obtained inductivity tensors of helical magnetic textures for various ${\bm q}$-directions.} The $\theta=0$ represents the ${\bm q}$-direction parallel to the $x$ axis. The solid symbols are the data obtained by the micromagnetic simulations, and the solid curves represent the corresponding trigonometric functions multiplied by $\lambda_1 = 3.11 \times 10^{-21}$ H~m. The simulations were done for $\beta = 0$.}
\end{figure}

The second category consists of tensors that have two inequivalent components, $\begin{pmatrix}
   \lambda_1 & 0 \\
   0 & \lambda_2
\end{pmatrix}$, and thus, off-diagonal components can appear if arbitrary Cartesian coordinates are chosen. For instance, the matrix $R(\theta)$ that rotates Cartesian coordinates clockwise by $\theta$ transforms the diagonalized tensor into a nondiagonal form:
\begin{eqnarray}
\begin{pmatrix}
   \lambda_1 & 0 \\
   0 & \lambda_2
\end{pmatrix}
\xrightarrow{R(\theta)}
\begin{pmatrix}
   \lambda_1\cos^2 \theta + \lambda_2 \sin^2 \theta & (\lambda_1-\lambda_2)\sin \theta \cos \theta \\
   (\lambda_1-\lambda_2)\sin \theta \cos \theta & \lambda_1\sin^2 \theta + \lambda_2\cos^2 \theta
\end{pmatrix}
.
\end{eqnarray}
Thus, $\tilde{L}_{xy} = \tilde{L}_{yx}$ can be either positive or negative depending on the selection of Cartesian coordinates. An example of this category is a single-${\bm q}$ helix, in which the inductance tensor is diagonalized, for instance, when the $x$-axis is chosen parallel to the helical-${\bm q}$ vector. An important feature of an ideal single-${\bm q}$ helix is that the local magnetic moments show no modulation along the direction perpendicular to ${\bm q}$. Hence, no STT effect is expected for the current along the $y$-axis, resulting in $\lambda_2 = 0$. Thus, for the case of an ideal single-${\bm q}$ helix, the diagonalized form and its orthogonal transformation are given as:
\begin{eqnarray}
\begin{pmatrix}
   1 & 0 \\
   0 & 0
\end{pmatrix}
\xrightarrow{R(\theta)}
\begin{pmatrix}
   \cos^2 \theta & \sin \theta \cos \theta \\
   \sin \theta \cos \theta & \sin^2 \theta
\end{pmatrix}
.
\end{eqnarray}
Figure 4 displays the comparison between the numerically obtained inductivity tensors of various ${\bm q}$-direction helices and the orthogonal transformation of $\lambda_1 = 3.11 \times 10^{-21}$ H~m and $\lambda_2 = 0$, which is an approximate inductivity tensor of the single-${\bm q}$ helix with $\theta = 0^{\circ}$ (Fig.~2a). Although the simulated single-${\bm q}$ helices are more or less affected by random disorder and the open boundaries of the system, the overall tendency is well reproduced by orthogonal transformation. This observation demonstrates that the above arguments based on orthogonal transformation of the symmetric inductivity tensor are helpful when considering a single-${\bm q}$ helix with arbitrary ${\bm q}$ direction.

For a more complicated magnetic texture, positive $\lambda_1$ and $\lambda_2$ with $\lambda_1 \neq \lambda_2$ may be expected. For instance, a mutidomain of single-$\bm{q}$-helices obviously belongs to this category. In contrast, a long-range ordered state belonging to this category may not be so clear. A candidate of this category is likely a magnetic texture that has multiple ${\bm q}$ vectors with different wavenumbers, such as those observed in EuAg$_4$As$_2$ \cite{EuAg4As2} and  EuAl$_4$ \cite{OrthoSk}; however, such an anisotropic magnetic texture is beyond the scope of our model Hamiltonian based on the continuum approximation [see Eq.~(\ref{Hamiltonian}) in the Methods section].

To conclude, we have revealed analytically and numerically the symmetry of the emergent inductance tensor exhibited by pinned magnetic textures. We focused on the adiabatic limit, where the inductance tensor is well defined by $e_i = \tilde{L}_{ik}\frac{{\rm d}j_k}{{\rm d}t}$. We thus found that the inductance tensor is a real symmetric tensor, and hence, the presence and magnitude of the transverse component are determined by the degree to which the measurement axis is tilted from the principal axis that generates the diagonalized tensor. As a natural consequence of the real symmetric tensor, the transverse component does not change sign with respect to the magnetic field reversal. These fundamental aspects of the emergent inductance tensor will be useful when exploring the transverse inductive response in a magnetic texture. However, it must also be noted that when nonadiabaticity is not negligible, the electric response produced by magnetic textures may not be described by an inductance tensor in the strict sense defined by Eq.~(\ref{tensorL}).

\section*{M\lowercase{ethods}}
\subsection*{Numerical model}
In this study, we consider long-period helical magnetic textures that are stabilized by the Dzyaloshinskii-Moriya (DM) interaction \cite{Dzyaloshinskii, Moriya}. We consider both a clean system without any disorder and dirty systems including randomly distributed disorder. Our model Hamiltonian is:
\begin{eqnarray}
\label{Hamiltonian}
\begin{split}
\mathscr{H} = &\int \frac{{\rm d}^3r}{a^3} \left[ \frac{J}{2}(\nabla\bm{m})^2 + D\bm{m}\cdot(\nabla\times\bm{m}) \right] \\
&- \sum_{k \in \Lambda} \int_{V_k} {\rm d}^3r \: K_{{\rm imp}} (\bm{m}_k \cdot \bm{n}_{{\rm imp}, k})^2
\end{split}
\end{eqnarray}
where $J$ is the Heisenberg exchange energy, $D$ is the DM interaction and $a$ is the lattice constant. The extrinsic pinning effect is controlled by the last term of Eq.~(\ref{Hamiltonian}), which is introduced to randomly selected cells to break the translational symmetry: $K_{{\rm imp}} (>0)$ represents the magnetic-easy-axis anisotropy along a randomly chosen direction, $\bm{n}_{{\rm imp}, k}$, at the $k$-th cell (the cell volume $V_k$ is $3^3$ nm$^3$), and $\Lambda$ is a set of random numbers. The disorder density displayed in Table I represents the ratio of the number of cells with finite $K_{{\rm imp}}$ to the total number of cells (243$\times$243).

When simulating the current-induced dynamics of a given helical magnetic structure, we insert the spin Hamiltonian into the following Landau-Lifshitz-Gilbert (LLG) equation \cite{LLG}:
\begin{eqnarray}
\label{LLG}
\begin{split}
\frac{{\rm d}\bm{m_r}(t)}{{\rm d}t}=-\frac{|\gamma|}{1+\alpha^2} \frac{{\rm d}\mathscr{H}}{{\rm d}\bm{m_r}} &\times \bm{m_r} - \frac{\alpha |\gamma|}{1+\alpha^2}\left[\bm{m_r}\times \left(\frac{{\rm d}\mathscr{H}}{{\rm d}\bm{m_r}} \times \bm{m_r} \right)\right]\\
+ \frac{1}{1 +\alpha^2}\{ (1&+\beta \alpha)\bm{m_r} \times [\bm{m_r}\times (\bm{u}\cdot \bm{\nabla})\bm{m_r}]\\
&+ (\beta - \alpha)[\bm{m_r}\times (\bm{u}\cdot \bm{\nabla})\bm{m_r}] \},
\end{split}
\end{eqnarray}
where $\bm{u}$ represents the spin drift velocity, $\alpha$ is the Gilbert damping constant, $\beta$ is a dimensionless constant that characterizes the nonadiabatic electron spin dynamics, and $\gamma$ $(>0)$ is the gyromagnetic ratio;
$\bm{u}$ is related to the electric current density $\bm{j}$ by $\bm{u} = \frac{P\mu_{\rm B}}{2|e|M_{\rm s}(1+\beta^2)}\bm{j}$, where $\mu_{\rm B}$ is the Bohr magneton and $M_{\rm s}$ is the saturation magnetization.
When implementing the micromagnetic simulation, we use the open software MuMax3 \cite{Mumax1, Mumax2}.
We choose the following parameter set:
$J/(2a^3) = 1.8\times10^{-11}$ J m$^{-1}$, $D/a^3 = 2.8\times10^{-3}$ J m$^{-2}$, $M_{\rm s} =2.45\times10^5$ A m$^{-1}$, $P = 1$, and $\alpha = 0.04$.

In the simulation, we apply a current density of a sufficiently small magnitude so that the magnetic system is certainly in the linear-response regime; that is, with respect to the input alternating electric current along the $x$ or $y$ direction, $j_i(t) = j_{0, i}\sin \omega t$ ($i = x, y$), the magnetic system is in the pinned regime, and the output AC emergent voltage, $V_{e, i}(t)$, is $\propto j_{0, k}\omega \cos \omega t$ $(i, k = x, y)$. Based on these observations, $L_{ij}$ is derived from the following equations:
\begin{eqnarray}
V_{e, i}(t) = \langle e_i(t) \rangle \ell = L_{ij}\frac{{\rm d}(I_j(t))}{{\rm d}t},
\end{eqnarray}
where $\langle \cdots \rangle$ denotes a spatially averaged value, the system length $\ell$ is $243 \times 3$ nm, and $I = jS$ with the cross-section area $S = 243 \times 1 \times 3^2$ nm$^2$. In the present frequency range ($\leq$ 100 MHz), it is confirmed that the inductivity is independent of $\omega$ (i.e., $\langle e_i \rangle \propto \omega$) (Fig.~S2) and the $\alpha$ dependence of the numerical results is negligibly small (Fig.~S3) (see also Supplementary Note 4). The numerical accuracy of MuMax3 is $\Delta{\bm m}/|{\bm m}| \sim 10^{-7}$, and the typical increment in one time step (4 ps) is $\sim$$10^{-5}$ under the current application of $\sim$10$^{10}$ A m$^{-2}$. This finite accuracy eventually gives rise to an uncertainty of $\sim$$10^{-23}$ H m in the calculated inductivity.

In the numerical simulation, a uniform current density is considered to understand fundamental aspects of the inductivity tensor. On the other hand, the local $\rho_{xx}$ and $\rho_{yx}$ may be non-uniform in real material, reflecting spatial variations in magnetic textures. Nevertheless, the uniform current is a good approximation as long as $\langle \rho_{xx} \rangle \gg \delta\rho_{xx}, \delta\rho_{yx}$, where $\delta\rho_{xx}$ and $\delta\rho_{yx}$ represent magnitude of the spatial variations. For instance, in the chiral magnet MnSi at 10 K, the presence or absence of the metastable skyrmion lattice changes $\rho_{xx}$ and $\rho_{yx}$ by $\approx$50 n$\Omega$ cm and $\approx$30 n$\Omega$ cm, respectively, whereas $\rho_{xx} \approx 5$ $\mu$$\Omega$ cm \cite{Oike}. Such magnetic-texture-dependent $\rho_{xx}$ and $\rho_{yx}$ imply that $\langle \rho_{xx} \rangle \gg \delta\rho_{xx}, \delta\rho_{yx}$ holds, although the precise estimation of the spatial variations is experimentally difficult; thus, the current uniformity is well expected. If $\delta\rho_{xx}$ and $\delta\rho_{yx}$ are significant, the current distribution should be determined self-consistently; for instance, see \cite{inhomo}.

\subsection*{Initial-state preparation}
To obtain various metastable magnetic textures, a pristine helical texture with a different oblique angle of the helical ${\bm q}$-vector, a random spin configuration, or an SkL is prepared as an initial state and then relaxed under zero current. Note that imposing the open-boundary condition and introducing impurity sites are key in obtaining the intended magnetic textures.

\section*{\uppercase{D}\lowercase{ata availability}}
The data used in this work are available from the corresponding author upon reasonable
request.

\section*{\uppercase{a}\lowercase{cknowledgments}}
	The authors thank N. Nagaosa and Y. Fujishiro for their valuable discussions. This work was partially supported by JSPS KAKENHI (Grants No.~20K03810, No.~18H05225, No.~23K03291 and No.~21H04442), JST CREST (Grants No.~JPMJCR1874 and No.~JPMJCR20T1).

\section*{\uppercase{c}\lowercase{ompeting interests}}
The authors declare no competing interests.

\section*{\uppercase{a}\lowercase{uthor contributions}}
S.F. conducted the calculations and analyzed the data. F.K. conceived the project and wrote the draft with S.F. and W.K. All the authors discussed the results and commented on the manuscript.

\renewcommand{\figurename}{Fig.~S}
\renewcommand{\tablename}{Table~S}
\setcounter{figure}{0}
\setcounter{table}{0}

\section*{S\lowercase{upplementary} N\lowercase{ote} 1: I\lowercase{mpact of nonadiabaticity}}
In the literature [16], it has been numerically demonstrated that when nonadiabaticity is considered (i.e., $\beta \neq 0$), the inductivity value defined using the emergent electric field under an AC current does not agree with that defined using the current-induced energy increase. This observation contradicts the conclusion derived from $e = \tilde{L}\frac{{\rm d}j}{{\rm d}t}$ and therefore implies that a simple equivalent circuit, such as a resistance ($R$)--inductance ($L$) series or $R$$\parallel$$L$ parallel circuit, can not consistently explain the electric and energetic responses of the system under the current. This fact also implies that the procedure for deriving the inductance from the observed impedance is not well defined. Therefore, the inductivity value of a magnetic texture is no longer well-defined when $\beta \neq 0$, at least in the present framework. For convenience, however, adopting a $R$--$L$ series circuit, one may define an effective inductivity in the low-frequency regime by $\tilde{L}^{\rm eff}_{ij} = \Im [\rho_{ij}(\omega)- \rho_{ij}(0)]/\omega$ and use this to describe the electric response. However, note that this effective inductivity or inductance ($L^{\rm eff}_{ij}= \tilde{L}^{\rm eff}_{ij}\frac{\ell}{S}$) no longer represents the genuine inductivity or inductance; i.e., $\tilde{L}^{\rm eff}_{ij} \neq \tilde{L}_{ij}$ Thus, there is no reason why the constraints associated with $\tilde{L}_{ij}$ [i.e., the diagonal components should be positive; the tensor should be symmetric; and $\tilde{L}_{xx}\tilde{L}_{yy} \geq (\tilde{L}_{xy})^2$] should also apply to $\tilde{L}^{\rm eff}_{ij}$. In fact, the diagonal $\tilde{L}^{\rm eff}_{ii}$ may be negative, whereas the diagonal $\tilde{L}_{ii}$ is invariably positive. For instance, when $R$--capacitance ($C$) parallel circuit is considered, $L^{\rm eff}_{xx}$ at low frequencies is negative, $-R^2C$, whereas the energy stored under a direct current (DC) is positive, $\frac{1}{2}(R^2C)I^2$. Thus, the effective inductivity or inductance contradicts the physics that $V = L\frac{{\rm d}j}{{\rm d}t}$ encompasses, and therefore, $L^{\rm eff}_{ij}$ should be viewed merely as a value that characterizes the electric response. The previous results [16] suggest that although $\tilde{L}^{\rm eff}_{ij} \neq \tilde{L}_{ij}$ in general, $\tilde{L}^{\rm eff}_{ij}$ becomes identical to $\tilde{L}_{ij}$ when $\beta=0$ (the adiabatic limit).

With this in mind, here we show how the effective inductivity $\tilde{L}^{\rm eff}_{ij}$ changes as $\beta$ increases. In this context, we note that several theoretical studies [42--45] have revealed that when $\beta$ is finite, the expression of the emergent electric field should be modified as follows:
\begin{align}
\label{BetaEEF}
    e_i &= \frac{P\hbar}{2|e|}\bm{m} \cdot (\partial_i\bm{m}\times\partial_{t}\bm{m}) - \beta\frac{P\hbar}{2|e|}(\partial_i\bm{m}\cdot\partial_{t}\bm{m}).  \tag{S1}
\end{align}
Thus, using this equation, we numerically obtain $e_i$ and thus $\tilde{L}^{\rm eff}_{ij}$ at low frequencies for a single-${\bm q}$ helix with $\theta = 0^{\circ}$, a maze helix, and a skyrmion lattice: the results are summarized in Supplementary Table SI-SIII, respectively.

In the single-${\bm q}$ helix with $\theta = 0^{\circ}$ (Table SI), finite $\beta$ leads exclusively to a decrease in $\tilde{L}^{\rm eff}_{xx}$, as discussed in the literature [16, 46]. For even larger $\beta$, $\tilde{L}^{\rm eff}_{xx}$ can be negative [46], but this conclusion relates to the effective inductivity and does not necessarily contradict the fact that the diagonal components of $\tilde{L}_{ij}$ must be positive. In contrast, no clear $\beta$ dependence is seen in the off-diagonal components of $\tilde{L}^{\rm eff}_{ij}$. The absence of the off-diagonal components will be corroborated from a symmetry point of view in Supplementary Note 3. The effective inductivity of a single-${\bm q}$ helix with arbitrary $\theta$ at finite $\beta$ is obtained immediately by applying the corresponding orthogonal transformation to the results of $\theta = 0^{\circ}$, and thus the results are omitted here. For the maze helix (Table SII), the effect of nonadiabaticity is weak.

The impact of nonadiabaticity on the skyrmion lattice (Table SIII) is different from that on the single-${\bm q}$ helix. We find that as seen in $\tilde{L}^{\rm eff}_{ij}(\beta)- \tilde{L}_{ij}$, the antisymmetric off-diagonal components become pronounced as $\beta$ increases. As demonstrated in the main text and Fig.~1, $e_i = \tilde{L}_{ik}\frac{{\rm d}j_k}{{\rm d}t}$ must result in a symmetric inductivity tensor, $\tilde{L}_{ik}=\tilde{L}_{ki}$. Thus, the emergence of the antisymmetric components in $\tilde{L}^{\rm eff}_{ij}(\beta)$ again indicates that at finite $\beta$, $\tilde{L}^{\rm eff}_{ij}$ is a different quantity from the inductivity tensor $\tilde{L}_{ij}$; only when $\beta = 0$ (the adiabatic limit), $\tilde{L}^{\rm eff}_{ij} = \tilde{L}_{ij}$, and hence, $\tilde{L}^{\rm eff}_{ij}$ is symmetric. The symmetry arguments on the SkL also support the emergence of the antisymmetric components in $\tilde{L}^{\rm eff}_{ij}$; for more details, see Supplementary Note 3.

\begin{table*}[h]
 \caption{Impact of nonadiabaticity on the effective inductivity tensor in the single-${\bm q}$ helix with $\theta = 0^{\circ}$. The unit is 10$^{-21}$ H~m.}
  \centering
 \begin{tabular}{ccc}
   \hline
   $\beta$ & \hspace{1cm} $\tilde{L}_{ij}$ or $\tilde{L}^{\rm eff}_{ij}(\beta)$ & \hspace{0.8cm} $\tilde{L}^{\rm eff}_{ij}(\beta)- \tilde{L}_{ij}$\\
   \hline \hline
  0 
& 
\hspace{1cm}
$\tilde{L}_{ij}=\tilde{L}^{\rm eff}_{ij}=$
$\begin{pmatrix}
   3.11 & -0.07 \\
   -0.07 & 0.07
\end{pmatrix}$
& 
\hspace{0.8cm}
\\
0.02
& 
\hspace{1cm}
$\tilde{L}^{\rm eff}_{ij}=$
$\begin{pmatrix}
   2.91 & -0.07 \\
   -0.06 & 0.07
\end{pmatrix}$
& 
\hspace{0.8cm}
$\begin{pmatrix}
   -0.20 & 0.00 \\
   0.01 & 0.00
\end{pmatrix}$
\\
0.04
& 
\hspace{1cm}
$\tilde{L}^{\rm eff}_{ij}=$
$\begin{pmatrix}
   2.32 & -0.04 \\
   -0.04 & 0.06
\end{pmatrix}$
& 
\hspace{0.8cm}
$\begin{pmatrix}
   -0.79 & 0.03 \\
   0.03 & -0.01
\end{pmatrix}$
\\
 \hline
  \end{tabular}
\end{table*}

\begin{table*}[h]
\caption{Impact of nonadiabaticity on the effective inductivity tensor in the maze helix. The unit is 10$^{-21}$ H~m.}
  \centering
  \begin{tabular}{ccc}
   \hline
   $\beta$ & \hspace{1cm} $\tilde{L}_{ij}$ or $\tilde{L}^{\rm eff}_{ij}(\beta)$ & \hspace{0.8cm} $\tilde{L}^{\rm eff}_{ij}(\beta)- \tilde{L}_{ij}$\\
   \hline \hline
  0 
& 
\hspace{1cm}
$\tilde{L}_{ij}=\tilde{L}^{\rm eff}_{ij}=$
$\begin{pmatrix}
   2.23 & -0.07 \\
   -0.08 & 2.67
\end{pmatrix}$
& 
\hspace{0.8cm}
\\
0.02
& 
\hspace{1cm}
$\tilde{L}^{\rm eff}_{ij}=$
$\begin{pmatrix}
   2.25 & -0.09 \\
   -0.05 & 2.66
\end{pmatrix}$
& 
\hspace{0.8cm}
$\begin{pmatrix}
   0.02 & -0.02 \\
   0.03 & -0.01
\end{pmatrix}$
\\
0.04
& 
\hspace{1cm}
$\tilde{L}^{\rm eff}_{ij}=$
$\begin{pmatrix}
   2.26 & -0.09 \\
   -0.05 & 2.63
\end{pmatrix}$
& 
\hspace{0.8cm}
$\begin{pmatrix}
   0.03 & -0.02 \\
   0.03 & -0.04
\end{pmatrix}$
\\
 \hline
  \end{tabular}
\end{table*}

\begin{table*}[h]
\caption{Impact of nonadiabaticity on the effective inductivity tensor in the skyrmion lattice at 0.3 T. The unit is 10$^{-21}$ H~m.}
  \centering
  \begin{tabular}{ccc}
   \hline
    $\beta$ & \hspace{1cm} $\tilde{L}_{ij}$ or $\tilde{L}^{\rm eff}_{ij}(\beta)$ & \hspace{0.8cm} $\tilde{L}^{\rm eff}_{ij}(\beta)- \tilde{L}_{ij}$\\
   \hline \hline
  0 
& 
\hspace{1cm}
$\tilde{L}_{ij}=\tilde{L}^{\rm eff}_{ij}=$
$\begin{pmatrix}
   7.40 & -0.47 \\
   -0.47 & 8.75
\end{pmatrix}$
& 
\hspace{0.8cm}
\\
0.02
& 
\hspace{1cm}
$\tilde{L}^{\rm eff}_{ij}=$
$\begin{pmatrix}
   7.39 & -0.92 \\
   -0.02 & 8.74
\end{pmatrix}$
& 
\hspace{0.8cm}
$\begin{pmatrix}
   -0.01 & -0.45 \\
   0.45 & -0.01
\end{pmatrix}$
\\
0.04
& 
\hspace{1cm}
$\tilde{L}^{\rm eff}_{ij}=$
$\begin{pmatrix}
   7.36 & -1.37 \\
   0.43 & 8.71
\end{pmatrix}$
& 
\hspace{0.8cm}
$\begin{pmatrix}
   -0.04 & -0.90 \\
   0.90 & -0.04
\end{pmatrix}$
\\
 \hline
  \end{tabular}
\end{table*}

\clearpage

\section*{S\lowercase{upplementary} N\lowercase{ote} 2: E\lowercase{ffective inductivity tensor under the time-reversal operation}}

When the nonadiabaticity $\beta$ is finite, antisymmetric components may appear in the effective inductivity tensor, $\tilde{L}_{ij}^{\rm eff}(\bm{B}, \{ \bm{m}(\bm{r}) \})$, in addition to symmetric components, as most clearly seen in Table SIII. For clarity, we decompose $\tilde{L}_{ij}^{\rm eff}(\bm{B}, \{ \bm{m}(\bm{r}) \})$ into symmetric and antisymmetric parts:
\begin{equation}
\tilde{L}_{ij}^{\rm eff}(\bm{B}, \{ \bm{m}(\bm{r}) \}) = \tilde{L}_{ij}^{\rm{eff}, S}(\bm{B}, \{ \bm{m}(\bm{r}) \}) + \tilde{L}_{ij}^{\rm{eff}, A}(\bm{B}, \{ \bm{m}(\bm{r}) \}),
\notag
\end{equation}
where $\tilde{L}_{ij}^{\rm{eff}, S}$ and $\tilde{L}_{ij}^{\rm{eff}, A}$ represent the symmetric and antisymmetric components, respectively. By definition, $\tilde{L}_{ij}^{\rm{eff}, S}$ and $\tilde{L}_{ij}^{\rm{eff}, A}$ satisfy
\begin{equation}
\begin{split}
\tilde{L}_{ij}^{\rm{eff}, S} &= \tilde{L}_{ji}^{\rm{eff}, S}, \\
\tilde{L}_{ij}^{\rm{eff}, A} &= -\tilde{L}_{ji}^{\rm{eff}, A}.
\end{split}
\notag
\end{equation}
On the other hand, as discussed in the main text, Onsager's reciprocal theorem concludes:
\begin{equation}
\tilde{L}_{ji}^{\rm{eff}}(\bm{B}, \{ \bm{m}(\bm{r}) \}) = \tilde{L}_{ij}^{\rm{eff}} (-\bm{B}, \{ -\bm{m}(\bm{r}) \}).
\notag
\end{equation}
By combining the two relations, one can thus conclude:
\begin{align}
\label{sym}
\tilde{L}_{ij}^{\rm{eff}, S}(\bm{B}, \{ \bm{m}(\bm{r}) \})  &= \tilde{L}_{ij}^{\rm{eff}, S}(-\bm{B}, \{ -\bm{m}(\bm{r}) \}), \tag{S2}
\\
\label{antisym}
\tilde{L}_{ij}^{\rm{eff}, A}(\bm{B}, \{ \bm{m}(\bm{r}) \})  &= -\tilde{L}_{ij}^{\rm{eff}, A}(-\bm{B}, \{ -\bm{m}(\bm{r}) \}). \tag{S3}
\end{align}
Thus, Onsager's reciprocal theorem concludes that the symmetric and antisymmetric components of $\tilde{L}_{ij}^{\rm{eff}}$ are even and odd under time-reversal, respectively.

This analytic conclusion is also confirmed numerically. Figure S1 shows the magnetic-field dependence of the transverse effective inductivity, $\tilde{L}_{xy}^{\rm{eff}}$ and $\tilde{L}_{yx}^{\rm{eff}}$, at a finite $\beta$, 0.04. For the single ${\bm q}$-helix with $\theta = 45^{\circ}$, the $\tilde{L}_{ij}^{\rm{eff}}$ is approximately symmetric ($\tilde{L}_{xy}^{\rm{eff}} = \tilde{L}_{yx}^{\rm{eff}}$) and satisfies Eq.~(\ref{sym}) under time-reversal. For the SkL, in contrast, the $\tilde{L}_{ij}^{\rm{eff}}$ consists of both symmetric and antisymmetric components, which satisfy Eqs.~(\ref{sym}) and (\ref{antisym}) under time-reversal, respectively.

\begin{figure}
\includegraphics{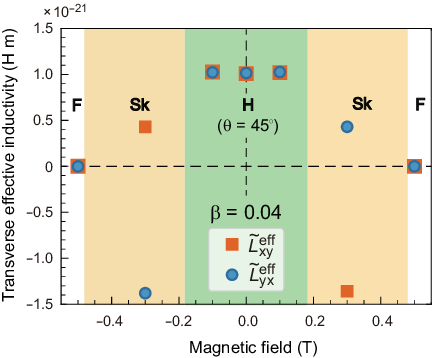}
\caption{\label{} \textbf{Magnetic-field dependence of the transverse effective inductivity, $\tilde{L}_{xy}^{\rm{eff}}$ and $\tilde{L}_{yx}^{\rm{eff}}$, at $\beta = 0.04$.} H, Sk, and F denote the single $\bm{q}$-helix with $\theta = 45^{\circ}$, skyrmion lattice, and ferromagnetic state, respectively. Note that the transverse components of $\tilde{L}_{ij}^{\rm{eff}}$ of the SkL includes both symmetric and antisymmetric components. The simulation parameters are the same as those used in Fig.~3 in the main text, except for $\beta$.}
\end{figure}

\section*{S\lowercase{upplementary} N\lowercase{ote} 3: S\lowercase{ymmetry arguments on a two-by-two polar tensor}}

In this section, we discuss how the symmetry of the magnetic texture places constrains on the allowed form of a two-by-two polar tensor. Note that such a symmetry argument holds whether or not the system is in the adiabatic limit ($\beta = 0$).

When the Cartesian coordinates are transformed by a clockwise rotation of angle $\theta$ with respect to the $z$ axis, the representation of an arbitrary two-by-two tensor,
$\begin{pmatrix}
   a & b \\
   c & d
\end{pmatrix}$,
changes accordingly. The new representation in the new Cartesian coordinates is given by
\begin{equation}
\begin{split}
&
\begin{pmatrix}
   \cos \theta & -\sin \theta \\
   \sin \theta & \cos \theta
\end{pmatrix}
\begin{pmatrix}
   a & b \\
   c & d
\end{pmatrix}
\begin{pmatrix}
  \cos \theta & -\sin \theta \\
  \sin \theta & \cos \theta
\end{pmatrix}
^{-1}
\\
&=
\begin{pmatrix}
\label{rotation}
   a \cos^2 \theta - b \sin \theta \cos \theta - c \sin \theta \cos \theta + d \sin^2 \theta & 
a \sin \theta \cos \theta + b \cos^2 \theta - c \sin^2 \theta - d\sin \theta \cos \theta  \\
   a \sin \theta \cos \theta - b \sin^2 \theta + c \cos^2 \theta - d\sin \theta \cos \theta & 
a \sin^2 \theta + b \sin \theta \cos \theta + c \sin \theta \cos \theta + d \cos^2 \theta. \end{pmatrix}
\end{split}
\tag{S4}
\end{equation}
If the rotation by $\theta$ is a symmetry operation of the system, the old and new representations should be identical (Neumann's principle). Similarly, when the Cartesian coordinates are transformed by a two-fold rotation with respect to the $x$ axis (denoted by $C_{2x}$), 
$\begin{pmatrix}
   1 & 0 \\
   0 & -1
\end{pmatrix}$,
the new representation of an arbitrary tensor is given by
\begin{equation}
\begin{split}
\begin{pmatrix}
   1 & 0 \\
   0 & -1
\end{pmatrix}
\begin{pmatrix}
   a & b \\
   c & d
\end{pmatrix}
\begin{pmatrix}
  1 & 0 \\
   0 & -1
\end{pmatrix}
^{-1}
=
\begin{pmatrix}
\label{C2x}
   a & -b \\
   -c & d
\end{pmatrix}
.
\end{split}
\tag{S5}
\end{equation}
If $C_{2x}$, which may be followed by translation, is a symmetry operation of the system, Neumann's principle concludes
\begin{equation}
\label{zero}
b=0, \hspace{1cm} c=0.
\tag{S6}
\end{equation}
The same conclusion is obtained when $C_{2y}$ (a two-fold rotation with respect to the $y$ axis), which may be followed by translation, is a symmetry operation of the system. Thus, the symmetry that the magnetic texture has places strong constraints on the allowed form of a two-by-two tensor. 

Below, we shall first discuss the symmetry constraints on the symmetric component of an arbitrary two-by-two tensor, 
$\begin{pmatrix}
   a & \frac{b+c}{2} \\
   \frac{b+c}{2} & d
\end{pmatrix}$.
When the system has, for instance, three-fold ($C_3$) symmetry $(\theta = \pm \frac{2\pi}{3})$, Neumann's principle leads to
\begin{equation}
\begin{pmatrix}
   a & \frac{b+c}{2} \\
   \frac{b+c}{2} & d
\end{pmatrix}
=
\begin{pmatrix}
   \frac{1}{4}a + \frac{\sqrt{3}}{2}\frac{b+c}{2} + \frac{3}{4}d & -\frac{\sqrt{3}}{4}a - \frac{1}{2}\frac{b+c}{2} + \frac{\sqrt{3}}{4}d \\
   -\frac{\sqrt{3}}{4}a - \frac{1}{2}\frac{b+c}{2} + \frac{\sqrt{3}}{4}d & \frac{3}{4}a - \frac{\sqrt{3}}{2}\frac{b+c}{2} + \frac{1}{4}d
\end{pmatrix}
.
\notag
\end{equation}
Solving this equation, one obtains
\begin{equation}
a=d, \hspace{1cm} \frac{b+c}{2}=0.
\notag
\end{equation}
The symmetry arguments thus conclude that a two-by-two symmetric tensor in the presence of $C_{3z}$ symmetry should be in the form of
$\begin{pmatrix}
   1 & 0 \\
   0 & 1
\end{pmatrix}$, 
regardless of whether the system has $C_{2x}$, which may be followed by translation, as a symmetry operation. The same conclusion is obtained when the system has four-, six-, or infinite-fold symmetry with respect to the $z$ axis ($C_{4z}, C_{6z}$ or $C_{\infty z}$). In contrast, when the system has only two-fold symmetry with respect to the $z$ axis ($C_{2z}$), Neumann's principle places no constraint because an arbitrary two-by-two tensor is invariant under $C_{2z}$ [see Eq.~(\ref{rotation})].

Next, we shall consider the antisymmetric component of an arbitrary two-by-two tensor, 
$\begin{pmatrix}
   0 & \frac{b-c}{2} \\
   -\frac{b-c}{2} & 0
\end{pmatrix}$.
As demonstrated in Table SIII, when nonadiabaticity is finite ($\beta \neq 0$), an antisymmetric component may appear in $\tilde{L}^{\rm eff}_{ij}$ (not in $\tilde{L}_{ij}$, of which derivation from the observed impedance is not well defined when $\beta$ is finite). Note that
$\begin{pmatrix}
   0 & 1 \\
   -1 & 0
\end{pmatrix}$
is invariant under arbitrary rotation with respect to the $z$ axis [Eq.~(\ref{rotation})], and hence, the rotational symmetry with respect to the $z$ axis does not place any constraint on the antisymmetric component. In contrast, when the system has $C_{2x}$ or $C_{2y}$, which may be followed by translation, a finite antisymmetric component is prohibited, as derived in Eqs.~(\ref{C2x}) and (\ref{zero}). 

The above symmetry consideration is consistent with the numerical observations on $\tilde{L}^{\rm eff}_{ij}$ of a single-$\bm{q}$ helix with $\bm{q}$$\parallel$$x$ (Table SI). The magnetic texture has $C_{2x}$ followed by translation as a symmetry operation, and hence, a finite antisymmetric component is prohibited, regardless of $\beta$. As discussed in the main text, by further considering the spin-transfer torque mechanism, one can conclude that a single-$\bm{q}$ helix with $\bm{q}$$\parallel$$x$ exhibits $\tilde{L}^{\rm eff}_{ij}$ in the form of 
$\begin{pmatrix}
   1 & 0 \\
   0 & 0
\end{pmatrix}$,
regardless of $\beta$. The $\tilde{L}^{\rm eff}_{ij}$ of a single-$\bm{q}$ helix with arbitrary $\bm{q}$-direction can be obtained by the orthogonal transformation of the $\tilde{L}^{\rm eff}_{ij}$ with $\bm{q}$$\parallel$$x$ [Eq.~(14) in the main text], and this transformation of the symmetric tensor does not add any antisymmetric component. One can thus conclude that $\tilde{L}^{\rm eff}_{ij}$ of a single-$\bm{q}$ helix should be symmetric, regardless of $\beta$ and the $\bm{q}$-direction. Nevertheless, the application of a magnetic field along the $z$ direction breaks $C_{2x}$ and $C_{2y}$ symmetry, and hence a magnetic-field-induced antisymmetric component is not prohibited, at least, by symmetry.

For a maze helix, the magnetic texture has a global approximate $C_{\infty z}$ symmetry, and hence, the symmetric part of the off-diagonal components should be zero, as discussed in the main text. Furthermore, because a single-$\bm {q}$ helix with $\theta$ is transformed into that with $-\theta$ by $C_{2x}$ and $C_{2y}$, a maze-helix texture, which is a collection of randomly oriented single-$\bm{q}$ helices, has global approximate $C_{2x}$ and $C_{2y}$ symmetries. Hence, the antisymmetric part should also be zero. In this way, $\tilde{L}^{\rm eff}_{ij}$ of a maze helix is expected to be also symmetric with no off-diagonal components, regardless of $\beta$. This symmetry argument is well consistent with the numerical observations shown in Table SII, given that the calculations are performed for a system of finite size. 
 
For the case of a skyrmion magnetic texture, the system has neither $C_{2x}$ nor $C_{2y}$ symmetry operations. Hence, there is no symmetry reason to prohibit a finite antisymmetric component. This symmetry argument is consistent with the numerical observations shown in Table SIII.

\vspace{1cm}

\section*{S\lowercase{upplementary} N\lowercase{ote} 4: F\lowercase{requency and $\alpha$ dependences of the inductivity tensor}}

In this section, we shall consider the frequency and $\alpha$ dependences of the inductivity tensor at $\beta = 0$. In this paper, we deal with the low-frequency regime in which the magnetic texture can sufficiently track the steady-state magnetic texture corresponding to each instantaneous current value. In other words, during AC application, the magnetic texture at a given instant is determined by the instantaneous current value at that instant, and it does not depend on the immediately preceding state. In fact, we found that below 100 MHz, the emergent electric field is proportional to $\omega$ (Fig.~S2), or equivalently, the inductivity tensor is independent of $\omega$. This observation confirms that our numerical results are in the low-frequency regime. The $\alpha$ dependence of the inductivity tensor in the low-frequency regime is therefore expected to be dictated by that of the steady-state magnetic texture.

To consider the $\alpha$ dependence of the steady-state magnetic texture, it is useful to refer to the LLG equation expressed in an equivalent form different from Eq.~(16) (for example, see [23]):
\begin{equation}
\frac{{\rm d}\bm{m_r}(t)}{{\rm d}t} = -|\gamma | \frac{{\rm d}\mathscr{H}}{{\rm d}\bm{m_r}} \times \bm{m_r} + \alpha \bm{m_r} \times \frac{{\rm d}\bm{m_r}}{{\rm d}t} - (\bm{u}\cdot \bm{\nabla})\bm{m_r} + \beta \bm{m_r} \times [(\bm{u}\cdot \bm{\nabla})\bm{m_r}].
\notag
\end{equation}
This expression implies that the steady-state configurations of magnetic moments, in which $\frac{{\rm d}\bm{m_r}(t)}{{\rm d}t} = \bm{0}$, do not depend on $\alpha$. Thus, the inductivity tensor in the low-frequency regime is expected to be also insensitive to $\alpha$. Figure S3 displays the inductivity tensor for the case of a single-$\bm{q}$ helix with $\theta = 20^{\circ}$ with $\alpha = 0.04$, 0.08 and 0.12. As expected, the results are the same within the numerical error, demonstrating that the choice of $\alpha$ does not make a difference in the inductivity tensor.

\begin{figure}
\includegraphics{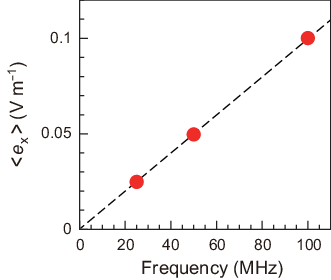}
\caption{\label{} \textbf{Frequency dependence of the amplitude of the emergent electric field.} The simulations were done for the single-$\bm{q}$ helix of $\theta=0^{\circ}$ at $\beta = 0$.}
\end{figure}

\begin{figure}
\includegraphics{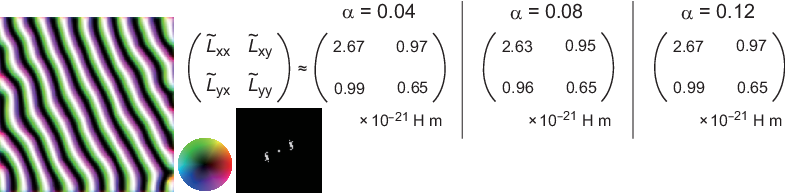}
\caption{\label{} \textbf{$\alpha$ dependence of the inductivity tensor.} The simulations were done for the single-$\bm{q}$ helix of $\theta=20^{\circ}$ at $\beta = 0$.}
\end{figure}

\end{document}